# Exceptional points of degeneracy and $\mathcal{PT}$-symmetry in photonic coupled chains of scatterers

Mohamed A. K. Othman[1], Vincenzo Galdi[2], and Filippo Capolino[1]

[1]*Department of Electrical Engineering and Computer Science, University of California, Irvine, Irvine, CA 92697, USA*

[2] *Waves Group, Department of Engineering, University of Sannio, I-82100 Benevento, Italy*

We demonstrate the existence of exceptional points of degeneracy (EPD) of periodic eigenstates in non-Hermitian coupled chains of dipolar scatterers. Guided modes supported by these structures can exhibit an EPD in their dispersion diagram at which two or more Bloch eigenstates coalesce, in both their eigenvectors and eigenvalues. We show a second-order modal EPD associated with the parity-time ($\mathcal{PT}$) symmetry condition, at which each particle pair in the double chain exhibits balanced gain and loss. Furthermore, we also demonstrate a fourth-order EPD occurring at the band edge. Such degeneracy condition was previously referred to as a *degenerate band edge* in lossless anisotropic photonic crystals. Here, we rigorously show it under the occurrence of gain and loss balance for a discrete guiding system. We identify a more general regime of gain and loss balance showing that $\mathcal{PT}$-symmetry is not necessary to realize EPDs. Furthermore, we investigate the degree of detuning of the EPD when the geometrical symmetry or balanced condition is broken. These findings open unprecedented avenues toward superior light localization and transport with application to high-$Q$ resonators utilized in sensors, filters, low-threshold switching and lasing.

## I. INTRODUCTION

Degeneracy in the state space of a dynamical system refers to points at which two, or more, physical eigenstates coalesce into one. This pervasive concept may give rise to interesting phenomena in many branches of physics [1–7]. In connection with electromagnetic (EM) waves, of particular interest for this study, it is well known that the propagation in closed guiding structures such as metallic waveguides or periodic structures, in the absence of energy dissipation or gain, is mathematically described in terms of a Hermitian operator [8]. This implies that the associated eigenspace is characterized by real-valued eigenvalues and it always constitutes a basis. As a consequence, with a few notable exceptions, strict modal degeneracy cannot take place, in the sense that eigenmodes associated with identical eigenvalues are generally linearly independent. However, certain degenerate conditions can be found where the system space is constructed from a generalized basis of eigenstates [1–3]. Simple examples of these degeneracies can be found in metallic waveguides at the cutoff or zero frequency. Another interesting example can be found at the transmission band edge of any periodic guiding structure, where there exists a *regular band edge* (RBE) at which forward and backward Bloch modes coalesce [9]. More pronounced degeneracy conditions, entailing the coalescence of three, four or more Bloch modes, can be found in a special class of anisotropic or birefringent photonic crystals. An example of a third-order degeneracy is found at the *stationary inflection point* (SIP) of a magnetic photonic crystal (MPhC) [10,11], whereas a fourth-order degeneracy is realized at a *degenerate band edge* (DBE) [12–17]. At such points of degeneracy, the group velocity vanishes and the local density of states exhibits a dramatic enhancement. These effects are demonstrated in lossless structures under RBE, SIP or DBE conditions [9,10,12,16]. In particular, the "frozen-wave" regime associated with the DBE condition [5,12,16,18–22] has been demonstrated to provide a better localization of light through large enhancement of the local density of states, as well as enhancement of gain in active configurations [16,22]. Moreover, several DBE implementations were carried out in coupled silicon waveguides [15,21] or 2D photonic crystals [23], with potential applications to lasers [14,16], and more recently at microwaves [24], for low-threshold oscillations [25–27] and efficient high power generation [22,25,28].

Recently, there has been a surge of interest in connection with degeneracies in system described by *non-Hermitian* operators. In these cases, the term "exceptional point" is used to indicate a non-Hermitian degeneracy where two or more eigenstates coalesce into one with the same *complex-valued* eigenvalue. Since the term "exceptional" may have different meanings in different disciplines, in what follows, we prefer to use the term "exceptional point of degeneracy" (EPD) so as to avoid possible ambiguities. The interest in this class of degeneracies is mainly motivated by their relevance in the study of parity-time- ($\mathcal{PT}$-) symmetric systems [2,3,29,30]. Originally introduced in quantum mechanics, as an alternative condition to ensure real-valued eigenspectra in the presence of pseudo-Hermitian [31–33] and non-Hermitian Hamiltonians, the $\mathcal{PT}$-symmetry concept has stimulated discussions in several branches of applied physics, including quantum field theories and quantum interactions [2,3,34,35]. Moreover, given the formal analogy with quantum mechanics, the $\mathcal{PT}$-symmetry concept has naturally been translated to paraxial optics [29,30,36,37]. In this case, practical implementations involve coupled waveguides and resonators exhibiting symmetric gain and loss distributions with suitable spatial modulation [30,37,38], although related effects were also demonstrated in passive lossy structures [39]. Aside from the interplay between gain and loss in coupled-mode structures supporting $\mathcal{PT}$-symmetry (see [30,40–42]), it is important to note that the $\mathcal{PT}$-symmetry is not a sufficient condition for a real-valued eigenspectrum. In fact, for a non-Hermiticity (i.e., gain/loss) level beyond a critical threshold, the system may encounter an EPD thereby undergoing a phase transition to a complex-valued eigenspectrum. This phenomenon is usually referred to "spontaneous symmetry breaking" [29,30,38]. In view of the comparatively simpler (with respect to quantum physics) implementations, optical $\mathcal{PT}$-symmetric structures have elicited a great deal of attention, leading to many interesting observations, including the demonstration of low-threshold lasing and laser absorbers [41,43–45], enhanced nonlinear effects [34,40-43], as well as metamaterial-based field manipulations [38,48,49]. In previous works, $\mathcal{PT}$-symmetry has been shown in discrete arrangements of resonators and also using the so-called "tight-binding" (TB) approach [48,50]. Moreover, EPDs have been also observed in 2D and 3D geometries [7,37,51].

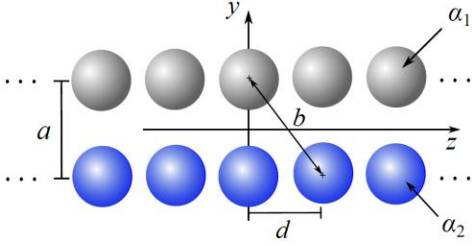

FIG. 1. Two coupled chains of polarizable particles with electrical dipolar polarizabilities $\alpha_1$ and $\alpha_2$. Supported Bloch modes polarized along the *x*-direction are investigated. The dispersion diagram of the modes guided along the *z*-direction may develop an exceptional point of degeneracy (EPD) under certain conditions.

In this paper, we study the emergence of EPDs in coupled chains of photonic scatterers exhibiting gain and loss. This configuration may constitute an interesting, and largely unexplored, photonic testbed for studying the properties of non-Hermitian systems. Moreover, it may find intriguing applications to light localization and transport.

Accordingly, the rest of the paper is laid out as follows. In Section II, we outline the problem statement. In Section III, we introduce the model utilized for the eigenmode analysis. In Section IV, we study the modal dispersion characteristics near second- and fourth-order EPDs, and elucidate the connections with the $\mathcal{PT}$-symmetry concept. Finally, in Section V, we provide some brief conclusions and discuss the implications and possible applications of our results.

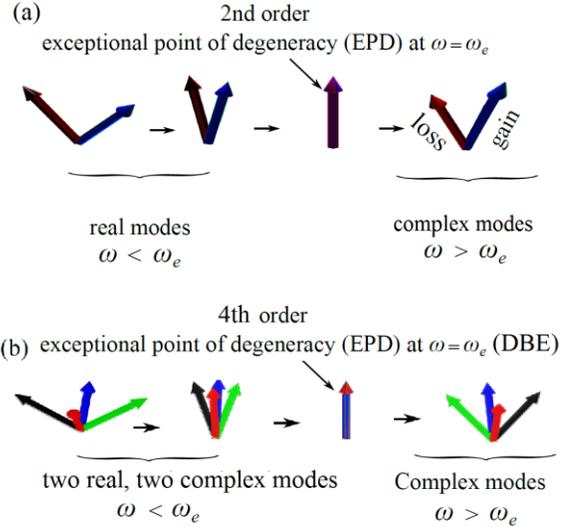

FIG. 2. Schematic representation (in three dimensional space) of the four-dimensional system state eigenvectors $\mathbf{\Psi}_l$ near an exceptional point of degeneracy. (a) Two eigenvectors coalesce at a second order EPD. (a) Four eigenvectors coalesce at a fourth order EPD, at the degenerate band edge.

## II. PROBLEM STATEMENT

As previously mentioned, we investigate the emergence of EPDs in coupled periodic chains of polarizable particles (see Fig. 1) exhibiting loss and gain. In particular, we derive a general (necessary and sufficient) condition for an EPD to occur, and we elucidate possible connections with the $\mathcal{PT}$-symmetry concept [29,30,39].

Our study yields two main results. First, we demonstrate the existence of EPDs by using a TB-based method. Such approach, based on the transfer-matrix method, is conventionally utilized in the study of photonic crystal waveguides [52,53], and has been previously employed to investigate the properties of non-linear magnetic resonators [48], discrete $\mathcal{PT}$-symmetric scatterers [50], and solitons in paired chains of dimers [50,51]. Second, we show the manifestation of both second- and fourth-order EPDs in such structures. Importantly, we demonstrate for the first time that DBEs (special kinds of EPDs) may exist in chains of discrete scatterers exhibiting loss and gain. We also elucidate the connection between these EPDs and the previously observed DBE effects in lossless structures. Our study exhibits several elements of novelty with respect to previous studies in the topical literature: By comparison with previous studies on chains of split rings [48], we emphasize that our proposed chain here is periodic along the *z*-direction, and we find degeneracies of different orders. Furthermore, although there are some previous studies of fourth-order EPDs in chains of multiple-resonators [6,54], here, we attain these effects in a photonic chain, composed of pairs of coupled scatterers (i.e., two coupled linear chains). Moreover, we also discuss some aspects that have been insofar overlooked, including the order of degeneracies, their relationship with DBE conditions explored in a very different contest [12,23], and the associated perturbation analyses. The evolution of the eigenstate vectors in the system, to be investigated thoroughly in Section IV, is illustrated schematically in Figs. 2(a) and (b) in the vicinity of second- and fourth-order EPDs, respectively. In our study, we will quantify the evolution of the complex Bloch wavenumbers in terms of frequency detuning and the gain/loss level, which will be rigorously defined in Section II and III.

We highlight that $\mathcal{PT}$-symmetry, also explored in our study, is a particular topology (and not the only one) available to realize a gain and loss balance scheme and construct real eigenspectra. Nevertheless, as shown hereafter, it is not a necessary condition for the existence of an EPD. In principle, polarizabilities with gain can be implemented in dielectric or plasmonic nanoparticles, with gain provided by fluorescent quantum dots cores [55] or dyes in the core or in the outer shell [56], for instance. We assume time-harmonic fields of the form $e^{-i\omega t}$, so that gain and loss correspond to complex-valued polarizabilities $\alpha = \alpha' + i\alpha''$, with $\alpha'' < 0$ and $\alpha'' > 0$, respectively.

## III. EIGENMODES IN TWO COUPLED CHAINS: TRANSFER MATRIX ANALYSIS

Referring to the schematic in Fig. 1, we consider a periodic chain of dipolar scatterer pairs in a homogenous medium. Each pair is characterized by two dipoles moments, with electric dipolar polarizabilities denoted by $\alpha_1$ and $\alpha_2$, separated by a distance $a$, and the chain's period is denoted by $d$. Accordingly, the chain's constitutive scatterers are located at $\mathbf{r}_{1,n} = (a/2)\hat{\mathbf{y}} + nd\hat{\mathbf{z}}$ and $\mathbf{r}_{2,n} = -(a/2)\hat{\mathbf{y}} + nd\hat{\mathbf{z}}$, respectively, with $n$ denoting an integer $n \in (-\infty, \infty)$. Here and henceforth, boldface symbols denote vector quantities and the caret denotes unit vectors. The polarizabilities relate the local electric field at the equivalent electric dipoles' locations to their moment, viz. $\mathbf{p}_{j,n} = \alpha_j \mathbf{E}^{\text{loc}}_{j,n}$, with $j = 1, 2$. The local electric field $\mathbf{E}^{\text{loc}}_{j,n}$ at $\mathbf{r}_{j,n}$ is produced by the infinite chain's dipole moments, in addition to any external excitation $\mathbf{E}^{\text{ext}}_{j,n}$, through the dyadic Green's function as

$$\mathbf{E}^{\text{loc}}_{j,n} = \mathbf{E}^{\text{ext}}_{j,n} + \sum_{\substack{p=1 \\ p \neq j}}^{2} \sum_{\substack{q=-\infty \\ q \neq n}}^{\infty} \underline{\underline{\mathbf{G}}}(\mathbf{r}_{j,n}, \mathbf{r}_{p,q}) \cdot \mathbf{p}_{p,q}, \quad (1)$$

where $\underline{\underline{\mathbf{G}}}(\mathbf{r}_{j,n}, \mathbf{r}_{p,q})$ is the electric-dipole dyadic Green's function (GF) [57,58]. By solving (1) in the absence of excitation ($\mathbf{E}^{\text{ext}}_{m,n} = 0$), we can compute the guided/leaky wave eigenmodes supported by the chain. In this study, we are only interested in the guided (bound) modes. An alternative representation of the fields can be accomplished via a combination of both spectral and spatial GFs, such as in the Ewald method for linear arrays [59–62].

Here, we make the following assumptions: (i) we consider a transverse polarization for which the excited dipole moments can be only oriented along the x-direction, so that $\mathbf{p}_{j,n} = p_{j,n}\hat{\mathbf{x}}$; (ii) we only consider interactions within nearest neighbors scatterers, justified by the fact that inter-particle distance is subwavelength. This approach, which resembles the TB formalism in solid state physics [52,53,63], was also utilized in [48,50,64,65] to analyze the general properties of discrete interactions in $\mathcal{PT}$-symmetric systems. Following these assumptions, we can recast (1) in a much simpler form:

$$\begin{aligned} p_{1,n} &= \alpha_1 \left[ G(d) p_{1,n+1} + G(d) p_{1,n-1} \right] + \\ &\quad + \alpha_2 \left[ G(b) p_{2,n+1} + G(a) p_{2,n} + G(b) p_{2,n-1} \right] \\ p_{2,n} &= \alpha_2 \left[ G(d) p_{2,n+1} + G(d) p_{2,n-1} \right] + \\ &\quad + \alpha_1 \left[ G(b) p_{1,n+1} + G(a) p_{1,n} + G(b) p_{1,n-1} \right] \end{aligned}, \quad (2)$$

where we have used the electric-dipole scalar GF $G(r) = G(\mathbf{r}, \mathbf{r}') = k^3 \exp(ikr)\left[(kr)^{-3} - i(kr)^{-2} - (kr)^{-1}\right]/C$ [58,66], with $C = -4\pi\varepsilon_0 \varepsilon_h$ (and $\varepsilon_h$ being the dielectric constant of the host medium), $r = |\mathbf{r} - \mathbf{r}'|$ and $b = \sqrt{a^2 + d^2}$ (see Fig. 1), and $k$ is the wavenumber in the host material. The equations in (2) can be cast in a form involving finite differences and a system evolution equation. However, our approach in this study relies on the construction of a transfer matrix that relates the dipole moments at two locations $\mathbf{r}_{m,n}$ and $\mathbf{r}_{j,n+1}$, from which we can calculate the band structure of the periodic chain. To this aim, it is expedient to define a four-dimensional *state vector* as $\mathbf{\Psi}(n) = \begin{bmatrix} p_{1,n} & p_{2,n} & p_{1,n-1} & p_{2,n-1} \end{bmatrix}^T$ (with the superscript "T" denoting the transpose) which describes the spatial evolution of the dipole moments in the coupled chains. It is important to stress that, even though the choice of the state vector is not unique [67], the eigenvalues of the system are invariant under any non-singular unitary (similarity) transformation of the state vector.

### A. State Vector Evolution and Transfer Matrix

Using (2), we construct a discrete matrix equation for the state vector evolution as

$$\mathbf{\Psi}(n+1) = \underline{\mathbf{T}}\mathbf{\Psi}(n), \quad (3)$$

where $\underline{\mathbf{T}}$ denotes the transfer matrix of the chain under the nearest neighborhood (i.e., TB) approximation. Such matrix can be written as

$$\underline{\mathbf{T}} = \underline{\mathbf{M}}\,\underline{\mathbf{V}}, \quad \underline{\mathbf{M}} = \begin{pmatrix} -\underline{\mathbf{1}} & \underline{\underline{\mathbf{A}}} \\ \underline{\underline{\mathbf{0}}} & -\underline{\underline{\mathbf{1}}} \end{pmatrix}, \quad \underline{\mathbf{V}} = \begin{pmatrix} \underline{\underline{\mathbf{0}}} & \underline{\underline{\mathbf{1}}} \\ -\underline{\underline{\mathbf{1}}} & \underline{\underline{\mathbf{0}}} \end{pmatrix}, \qquad (4)$$

where $\underline{\underline{\mathbf{1}}}$ is a 2×2 identity matrix, and $\underline{\underline{\mathbf{A}}}$ is a 2×2 matrix given by

$$\underline{\underline{\mathbf{A}}} = \begin{pmatrix} \dfrac{\alpha_2 G(d) + \alpha_1 \alpha_2 G(a) G(b)}{\alpha_1 \alpha_2 \left(G^2(d) - G^2(b)\right)} & \dfrac{-\alpha_2 G(b) - \alpha_2^2 G(a) G(d)}{\alpha_1 \alpha_2 \left(G^2(d) - G^2(b)\right)} \\ \dfrac{-\alpha_1 G(b) - \alpha_1^2 G(a) G(d)}{\alpha_1 \alpha_2 \left(G^2(d) - G^2(b)\right)} & \dfrac{\alpha_1 G(d) + \alpha_1 \alpha_2 G(a) G(b)}{\alpha_1 \alpha_2 \left(G^2(d) - G^2(b)\right)} \end{pmatrix}. \quad (5)$$

The transfer matrix $\underline{\mathbf{T}}$, as in the context of layered media analysis [8,68], obeys some fundamental properties, such as $\det(\underline{\mathbf{T}}) = 1$ [9,69,70]; other spectral properties will be further discussed hereafter. We seek Bloch-type wave (periodic) solutions of (3) in the form

$$\mathbf{\Psi}(n+1) = \zeta\,\mathbf{\Psi}(n), \qquad \zeta \equiv e^{ik_z d}, \qquad (6)$$

where $k_z$ is a generally complex-valued Bloch wavenumber of the guided mode supported by the chain, with the sign of the real and imaginary part determining the forward/backward, and propagating/evanescent character, respectively in a lossless structure. In the presence of gain and/or loss, these sign specifications may be violated in general. Note that a purely real-valued $k_z$ means that power is conserved for that mode [64], and this may occur in chains with balanced gain and loss, as we show hereafter (see Section IV below). We emphasize that $e^{ik_z d}$ is the eigenvalue of (6), not the Bloch wavenumber $k_z$; however it is natural to investigate the characteristics of $k_z$ since it allows for a straightforward assessment of the gain and loss balance [71]. Bloch eigenmodes that satisfy (3) and (6) are derived from the eigenvalue problem,

$$\underline{\mathbf{T}}\,\mathbf{\Psi}_l(n) = \zeta_l\,\mathbf{\Psi}_l(n), \qquad (7)$$

where $\mathbf{\Psi}_l(n)$ is the $l^{\text{th}}$ state eigenvector, with $l=1,2,3,4$. This yields four eigenvalues and corresponding eigenvectors. Note that the homogenous solutions of (3) are constructed from the four eigenvectors in (7) in the case where the matrix $\underline{\mathbf{T}}$ can be diagonalized. When $\underline{\mathbf{T}}$ is not diagonalizable, i.e., at an EPD, generalized eigenvectors are used instead of the regular eigenvectors in (7) [22]. The four eigenvalues of (7) are determined from

$$\det\left[2\cos(k_z d)\underline{\underline{\mathbf{1}}} - \underline{\underline{\mathbf{A}}}\right] = 0, \qquad (8)$$

which is further simplified to the transcendental form

$$4\cos^2(k_z d) - 2\cos(k_z d)\operatorname{Tr}(\underline{\underline{\mathbf{A}}}) + \det(\underline{\underline{\mathbf{A}}}) = 0. \qquad (9)$$

Note that from (9) we infer the symmetry property that both $k_z$ and $-k_z$ are solutions, as expected in view of the time-inversion symmetry that is still valid under the small-signal and linear-gain assumptions. Depending on whether an EDP occurs or not, the number of independent eigenvectors that satisfy (7) may vary from one to four. Indeed, though not always possible in general, the transfer matrix $\underline{\mathbf{T}}$ may be diagonalized so that,

$$\underline{\mathbf{T}} = \underline{\mathbf{U}}\,\underline{\mathbf{\Lambda}}\,\underline{\mathbf{U}}^{-1}, \qquad (10)$$

where $\underline{\mathbf{\Lambda}}$ is a diagonal matrix, whose entries are the eigenvalues of (6), and $\underline{\mathbf{U}}$ implements a similarity transformation. In this case, there would be four independent system state eigenvectors.

### B. Exceptional Points of Degeneracy (EPDs)

We now investigate a particular aspect of the dispersion diagram, namely, the emergence of EPDs. At an EPD, the matrix $\underline{\mathbf{U}}$ is singular, i.e., $\det[\underline{\mathbf{U}}] = 0$. Owing to the reciprocity (T-inversion symmetry) restriction of the system, we can only attain two different kinds of degeneracies: (i) a second-order degeneracy at which two eigenstates coalesce, with a multiplicity $m = 2$ of the eigenstates; and (ii) a fourth order degeneracy at which all eigenstates coalesce, with $m = 4$.

Accordingly, we investigate these two conditions in which the transfer matrix becomes similar to a matrix having Jordan blocks [72]. Under these conditions, a reduced number of regular eigenvectors will be found. In particular, when a fourth-order EPD occurs, (7) will possess one eigenvalue with multiplicity of four and only one regular eigenvector.

i) *Second order EPD*. At a second-order EPD, the transfer matrix is written as

$$\underline{\mathbf{T}} = \underline{\mathbf{W}}\begin{bmatrix} \underline{\underline{\mathbf{\Lambda}}}^+ & 0 \\ 0 & \underline{\underline{\mathbf{\Lambda}}}^- \end{bmatrix}\underline{\mathbf{W}}^{-1}, \\ \underline{\underline{\mathbf{\Lambda}}}^+ = \begin{pmatrix} \zeta_e & 1 \\ 0 & \zeta_e \end{pmatrix}, \quad \underline{\underline{\mathbf{\Lambda}}}^- = \begin{pmatrix} 1/\zeta_e & 1 \\ 0 & 1/\zeta_e \end{pmatrix}, \qquad (11)$$

where $\zeta_e = e^{ik_e d}$ is the EPD real eigenvalue of (7). The second-order EPD condition is found at an angular frequency $\omega = \omega_e$, and such degeneracy in the fundamental Brillouin zone $k_z \in [0, 2\pi/d]$ occurs between two Bloch modes having $k_z \in [0, \pi/d]$ (denoted by the superscript "+"); the other two modes with $k_z \in [\pi/d, 2\pi/d]$ (denoted by the superscript "–") must also coalesce in view of the symmetry conditions of the eigenvalues solutions in (7). Here, $\underline{\underline{\Lambda}}^+$ is a 2×2 Jordan block and $\underline{\underline{W}}$ is constructed from two regular and two generalized basis-eigenvectors. At $\omega = \omega_e$, homogenous solutions for the state vector in (3) are given in terms of two periodic (Bloch) modes having regular eigenvectors propagating as $e^{\pm i k_e n d}$, and two diverging solutions constructed from generalized eigenvectors that linearly grow as $n d e^{\pm i k_e n d}$. It is important to point out that, near the second-order EPD, the wavenumber $k_z$ can be written as a small perturbation of the ideal degeneracy condition with $k_z = k_e$, in terms of a fractional power expansion as

$$k_{z,l}(\omega) \cong (-1)^l k_e + h_l \delta^{1/2} + g_l \delta + \cdots, \qquad (12)$$

where $h_l$ and $g_l$ are the fractional series expansion coefficients for the four modes with $l=\{1,2,3,4\}$, and $\delta$ is a small perturbation parameter about the EPD. Such perturbation parameter identifies the detuning from the ideal EPD condition in the spectral evolution of the states, which could be observed via frequency detuning, gain and loss imbalance, or asymmetry in the chain (or in any other structural parameter). We recall that the perturbation analysis of degenerate or "defective" operators requires to deal with fractional power expansion, contrary to systems having only eigenvalue degeneracies (i.e., only coincident eigenvalues, but still a complete basis of eigenvectors [2,4,70,73]). As such, the fractional power series (12), also known as Puiseux series, is a direct consequence of the Jordan Block similarity [1,12,73]. Note that the principal root of $\delta$ is taken in (12). In Section III, we show the effect of two perturbation parameters (frequency detuning and gain/loss imbalance or asymmetry in the chain) separately, and their consequences on $\mathcal{PT}$-symmetry and the second-order EPD.

ii) *Fourth order EPD.* At a fourth-order EPD, the transfer matrix becomes similar to a four-dimensional Jordan matrix,

$$\underline{\underline{T}} = \underline{\underline{S}}\,\underline{\underline{\Lambda}}\,\underline{\underline{S}}^{-1}, \qquad \underline{\underline{\Lambda}} = \begin{pmatrix} \zeta_e & 1 & 0 & 0 \\ 0 & \zeta_e & 1 & 0 \\ 0 & 0 & \zeta_e & 1 \\ 0 & 0 & 0 & \zeta_e \end{pmatrix}, \qquad (13)$$

thereby implying a fourth-order degeneracy between all Bloch modes in the 4×4 system. Here, $\underline{\underline{\Lambda}}$ is a 4×4 Jordan matrix, and $\underline{\underline{S}}$ is constructed from one regular and three generalized basis-eigenvectors. In this particular case, we find that $\zeta_e = -1$. At $\omega = \omega_e$, homogenous solutions for the state vector in (3) at the fourth-order EPD are given in terms of one Bloch periodic mode having a regular eigenvector in (7) that propagates as $e^{i k_e n d}$, and three non-Bloch (non-periodic) diverging solutions constructed from a generalized set of eigenvectors growing as $n d e^{i k_e n d}$, $(n d)^2 e^{i k_e n d}$, and $(n d)^3 e^{i k_e n d}$ [22,70]. Similar to the second-order EPD, the wavenumber near a fourth-order EPD asymptotically follows the fractional power expansion

$$k_{z,l}(\omega) \cong k_e + h_l \delta^{1/4} + g_l \delta^{2/4} + \cdots, \qquad (14)$$

where $h_l$ and $g_l$ are the fractional series expansion coefficients for the $n$th eigenmodes, and $\delta$ is the perturbation factor.

In what follows, we quantitatively investigate the modal dispersion characteristics near a second- and fourth-order EPD of a chain composed of a pair of dipolar scatterers with gain and loss.

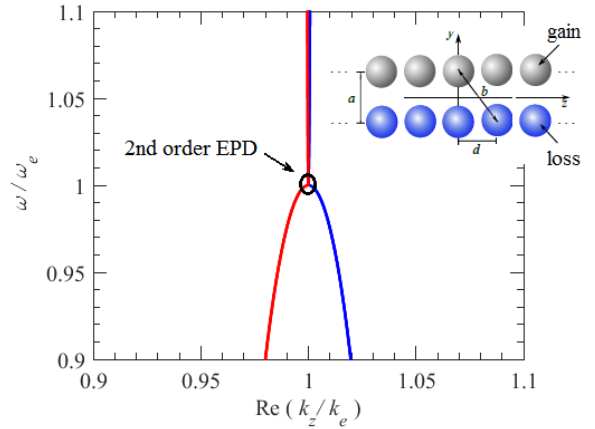

FIG. 3. Dispersion diagram of the two modes with positive $\mathrm{Re}(k_z)$ exhibiting an exceptional point of degeneracy at $\omega_e$. The $\mathcal{PT}$-symmetry allows for real modes when gain/loss balance and topological symmetry condition (19) is satisfied below the EPD $\omega < \omega_e$. Here $d=a=100$ nm, $\omega_e d/c = 0.02$ and $k_e = \pi/(5d)$, and the polarizabilities of the chain are found in Table I.

## IV. EPD AND $\mathcal{PT}$-SYMMETRY IN GAIN- AND LOSS-BALANCED COUPLED CHAIN

As mentioned in the previous section, degenerate states, if they exist, are characterized not only by the multiplicity of the eigenvalues in (7) but also by their geometric multiplicity, i.e., the linear dependence of the eigenvectors [1,3,6,72,74]. We assume that the wavenumber of the degenerate state is denoted by $\pm k_e$. Therefore, by invoking the eigenvalue multiplicity and symmetry conditions, the dispersion relation in (9) at the EPD takes the form

$$\left[\cos(k_z d) - \cos(k_e d)\right]^2 = 0. \quad (15)$$

Moreover, when an EPD occurs, the transfer matrix $\underline{\underline{\mathbf{T}}}$ in (4) can only be written in terms of Jordan blocks, and not in terms of diagonalized matrices. To derive the conditions on the polarizabilies of the coupled chain's scatterers in order for an EPD to occur, we compare (9) with (15), and obtain two conditions on the characteristic matrix $\underline{\underline{\mathbf{A}}}$, viz.,

$$\mathrm{Tr}(\underline{\underline{\mathbf{A}}}) = 4\cos(k_e d), \quad \text{and} \quad \det(\underline{\underline{\mathbf{A}}}) = 4\cos^2(k_e d). \quad (16)$$

Equations (16), along with (13), impose the following conditions:

$$\alpha_1^e \alpha_2^e = \frac{1}{4\cos^2(k_e d)\left(G^2(d) - G^2(b)\right) + G^2(a)} \equiv \xi,$$

$$\left(\alpha_1^e + \alpha_2^e\right) = \frac{-2G(a)G(b) + 4\left(G^2(d) - G^2(b)\right)\cos(k_e d)}{\left[4\cos^2(k_e d)\left(G^2(d) - G^2(b)\right) + G^2(a)\right]G(d)} \equiv \chi$$

(17)

where $\alpha_1^e \equiv \alpha_1(\omega_e)$ and $\alpha_2^e \equiv \alpha_2(\omega_e)$ are the required values for the polarizability to achieve a second or fourth-order EPD at a wavenumber $k_e$ and angular frequency $\omega_e$. Another necessary condition, besides (15), is that $\det[\underline{\mathbf{U}}] = 0$, which is implicitly satisfied from (14) through the constraint $\mathrm{Tr}^2(\underline{\underline{\mathbf{A}}}) = 4\det(\underline{\underline{\mathbf{A}}})$. Accordingly, the conditions in (17) on the polarizabilities are necessary and sufficient to attain the required EPD. The polarizabilities are obtained as solutions of (17) in terms of $\xi = \alpha_1^e \alpha_2^e$ and $\chi = \alpha_1^e + \alpha_2^e$ as

$$\alpha_{1,2}^e = \pm \frac{\chi}{2} \mp \sqrt{\chi^2 - 2\xi}, \quad (18)$$

for an EPD occurring at an angular frequency $\omega_e$. We highlight that some trivial conditions exist for the chain to develop an EPD, such as at zero frequency. In what follows, we focus on *non-trivial* EPDs, namely second- and fourth-order, in the presence of both gain and loss. We refer to "gain and loss balance" as the universal condition that guarantees the existence of an EPD in the spectrum of a coupled system described by non-Hermitian evolution equations as discussed in [72]. Indeed, $\mathcal{PT}$-symmetry is not a necessary condition for developing an EPD as shown in [72,75]. The more general balance condition is revealed in the chains when (18) is satisfied, resulting in an EPD in the chains spectra. Within this framework, $\mathcal{PT}$-symmetry is a special case, which would also lead to observing an EPD as discussed in the following.

### A. Second Order EPD and $\mathcal{PT}$-Symmetry

A second-order degeneracy indicates that the solutions $\zeta_{e,l}$ of the system (7) can take the values $\zeta_{e,1} \equiv \zeta_e \equiv e^{ik_e d}$ and $\zeta_{e,2} \equiv 1/\zeta_e \equiv e^{-ik_e d}$, with $k_e d \neq \pi$, i.e., *away* from the center of the Brillouin zone (defined here as the interval $k_z \in [0, 2\pi/d]$). Indeed, $k_z = \pi/d$ is the center of Brillouin zone where modes naturally coalesce, and it is well known to be a point where the group velocity vanishes if a bandgap exists [8,70]. To gain some physical insight into the conditions above, it is important to explore how the polarizabilities of the chain are constrained for an EP to occur. In what follows, we investigate different regimes of operations based on quasi-static approximations, and the effect of phase retardation on the EPD conditions.

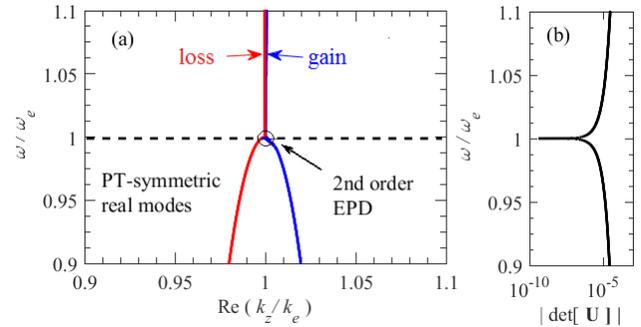

FIG. 4. (a) Same as Fig. 3. (b) Magnitude of the determinant of matrix $\underline{\mathbf{U}}$ that brings the transfer matrix into a diagonal form in (10).

*i) Quasi-static limit:* This applies when $kr \to 0$, with and $r$ being an arbitrary observation distance. Consequently, the GF follows its electrostatic limit $G(r) \to 1/(Cr^3)$. It is straightforward to see that, under such condition, an EPD can occur provided that $\left(\alpha_1^e + \alpha_2^e\right)^2 < 4\alpha_1^e \alpha_2^e$, since $G(d) > G(b)$ (the GF is real under this limiting case). Therefore, by imposing the conditions for an EPD to occur [(16) and (17)], we obtain the conjugate symmetry condition

$$\alpha_1^e = \left(\alpha_2^e\right)^*. \quad (19)$$

This condition implies that, when such low-frequency EPD occurs, one chain exhibits losses and the other exhibits gain that precisely compensates for the losses. This condition is inherently tied with the aforementioned $\mathcal{PT}$-symmetry concept. As typical in $\mathcal{PT}$-symmetric systems, the EPD is related to the *spontaneous symmetry breaking* phenomenon, and it constitutes the boundary that separates the "exact" and "broken" phases characterized by

real- and complex-valued eigenspectra, respectively. Such condition was rigorously satisfied in uniform coupled-waveguides [29,30,72,76], and here we showed that it holds in connection with periodic coupled chains of scatterers as well. Effects of field-retardation corrections in the GF are discussed next.

*ii)    Effect of GF phase retardation*

The $\mathcal{PT}$-symmetry with perfectly symmetric gain and loss balance governed by (19) is relevant when $kr \to 0$, corresponding to the quasi-static case described in the previous sub-section. However, when phase propagation is included in the GF, i.e., $kr$ assumes finite values, radiation losses exist due a non-vanishing imaginary part of the GF [57]. Therefore, at an EPD one expects that the conjugate-symmetry condition (19) is no longer rigorously satisfied due to the extra radiation (scattering) losses in the chain. Nevertheless, as we show below, an EPD can still occur since a gain loss *balance* can be achieved from the condition in (18). On the other hand, if one chooses the perfect conjugate *symmetry* condition (19) on the chain, an ideal EPD (where the eigenvectors are rigorously degenerate) can no longer be identified.

TABLE I. REQUIRED CHAIN POLARIZABILITIES TO REALIZE A SECOND ORDER EPD AT DIFFERENT NORMALIZED FREQUENCIES (UP TO 6 SIGNIFICANT DIGITS).

| $\omega_e d / c$ | $\alpha_1^e \times 10^{-32}$ [Cm$^2$V$^{-2}$] | $\alpha_2^e \times 10^{-32}$ [Cm$^2$V$^{-2}$] |
|---|---|---|
| 0.02 | $-3.59259 - i4.97264$ | $-3.59260 + i4.97256$ |
| 0.1  | $-3.61775 - i4.99120$ | $-3.6193 + i4.98184$ |
| 0.5  | $-3.93791 - i6.06210$ | $-4.18713 + i4.740$ |

We consider an example of a chain in vacuum (i.e., $\varepsilon_h = 1$) with $a = d = 100$ nm, and we select the EPD wavenumber to be $k_e d = \pi/5$, with $\omega_e d/c = 0.02$. This frequency implies that the period $d \approx 0.003\lambda_e$ (where $\lambda_e = 2\pi c/\omega_e$ is the wavelength in vacuum) is deeply subwavelength, thereby justifying the low-frequency assumption. By assuming the polarizabilities as frequency-independent in the vicinity of $\omega_e$, we obtain the values $\alpha_1 \cong \alpha_2^* = (-3.593 - 4.973) \times 10^{-32}$ [Cm$^2$V$^{-2}$] (see table I), in order to attain a second-order EPD at $\omega_e d/c = 0.02$. These values approximately satisfy (19) because of the low-frequency choice for this EPD to occur. In Fig. 3, we show the dispersion relationship of the two modes exhibiting the EPD. We only show the positive real part of the complex wavenumber within the region $\text{Re}(k_z) \in [0, \pi/d]$, but we stress that the wavenumber branches satisfying $\text{Re}(k_z) \in [\pi/d, 2\pi/d]$ also exhibit the EPD thanks to reciprocity. In Fig. 4, we also show the determinant (magnitude) of the similarity matrix $\underline{\mathbf{U}}$, which represents a quantitative metric of the closeness to an EPD condition. Indeed, at $\omega = \omega_e$, we observe that $|\det(\underline{\mathbf{U}})| \to 0$ indicating that the system eigenvector coalesce and cease to form a complete basis set [1,2,30]. As previously explained, exactly at the EPD there is no similarity transformation that diagonalizes the transfer matrix $\underline{\mathbf{T}}$, which in turn becomes similar to a matrix with Jordan blocks as in (11). In Fig. 5, we show the modes evolution in the complex $k_z$-plane [Re($k_z$)−Im($k_z$) plane] as the frequency increases. As it can be observed, the modal wavenumbers are almost real ( $|\text{Im}(k_z)/\text{Re}(k_z)| < 10^{-3}$ ) for $\omega < \omega_e$, and become almost complex conjugate pairs for $\omega > \omega_e$. Modes with purely real $k_z$ for $\omega < \omega_e$ are a fundamental consequence of the perfect gain-loss balance and symmetry of the system obeying an "exact" phase of $\mathcal{PT}$-symmetry, for large wavelength. Conversely, for $\omega > \omega_e$ such properties are violated, even though the gain and loss conjugate symmetry in (19) is satisfied, and the system enters the "broken" phase. By setting $\delta \equiv (\omega - \omega_e)$ in (12), and limiting the fractional power series to the first-order term, the asymptotic dispersion relationship near such second-order EPD is given by

$$\left(k_{z,l} - (-1)^l k_e\right)^2 \cong h_l^2 \delta, \text{ where } h_l^2 = 2\partial^2 k_{z,l}/\partial \omega^2 \text{ at } \omega = \omega_e.$$

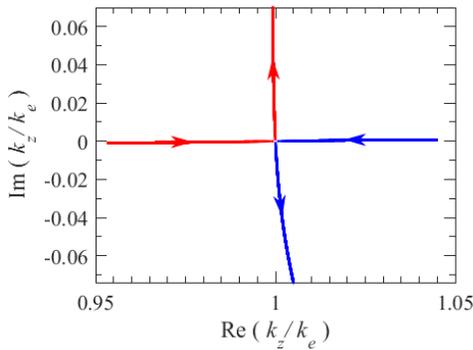

FIG. 5. Complex modal $k_z$ trajectory in the [Re($k_z$)-Im($k_z$)] plane varying as a function of frequency showing the modes coalescing at the second order EPD at $k_z = k_e$. Arrows show increasing frequency. Here, we used the same chain's parameters as in Fig. 3.

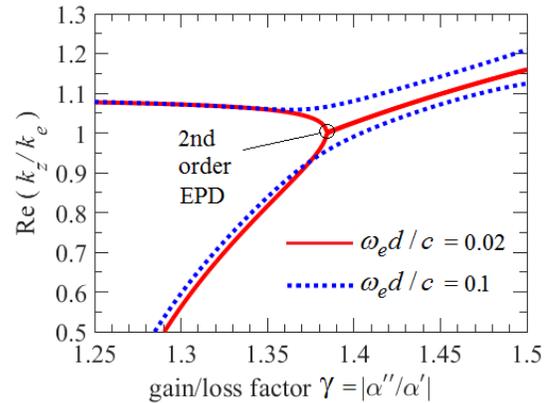

FIG. 6. Positive branch of Re[$k_z$] varying as a function of the gain/loss parameter $\gamma = |\alpha''/\alpha'|$, demonstrating the detuning near a 2nd order EPD for two cases of the normalized frequency $\omega_e d/c$. Chain parameters are as in Fig. 9. Note that the perfect gain/loss symmetry condition does not provide a clear EPD for the case with higher frequency.

We also investigate the detuning of the second-order EPD by varying the gain and loss values implemented in the polarizabilities from their optimal condition (18). For that purpose, we assume an exact gain and loss symmetry in the chain, i.e., $\alpha_{1,2} = \alpha'(1 \pm i\gamma)$, with the gain and loss normalized factor $\gamma$ defined as $\gamma = |\alpha''/\alpha'|$. We are interested in exploring how the mode characteristics change by varying the gain and loss factor $\gamma$. In Fig. 6, we show the positive branches of $k_z$ for two different normalized EPD frequencies $\omega_e d/c$, varying as a function of $\gamma$ for a chain with parameters $d=a=100$ nm, and we choose $\alpha' = \text{Re}(\alpha_1^e)$ from Table I for each value of $\omega_e d/c$. We clearly observe the occurrence of the EPD for the smaller electrical period ($\omega_e d/c = 0.02$). Moreover, the exact $\mathcal{PT}$-symmetric phase is observed at that frequency for a gain/loss parameter $\gamma$ less than a critical value ($\gamma \simeq 1.358$). At $\gamma \simeq 1.358$, the system undergoes spontaneous $\mathcal{PT}$-symmetry breaking, designating the EPD, and beyond this threshold the modes cease to be real. However, for the higher-frequency case of $\omega_e d/c = 0.1$, the EPD can no longer be attained with the exact gain and loss symmetry. Instead, as discussed previously, an asymmetry must be introduced due to radiation losses. In this connection, the reader is also referred to [72] in which different figure of merits were proposed in order to assess the *quality* or *evidence* of such EPD subject to perturbation due to disorders and imperfect gain and loss balance. To further elucidate this aspect, we also show the dispersion relationships of the modes belonging to a chain in which the polarizabilities are obtained from the symmetry design equations (19) to exhibit a second order EPD. We consider two frequencies for which $\omega_e d/c = 0.1$ and $\omega_e d/c = 0.5$ (i.e., increasing the frequency or period with respect to the case shown in Fig. 4) in Figs. 7 and 8, respectively. The corresponding polarizabilities of the chain that are evaluated from (18) to realize the EPD conditions at the above mentioned frequencies are

$\alpha_1 = (-3.61 - i4.99) \times 10^{-32}$ [$\text{C}\,\text{m}^2\text{V}^{-2}$],

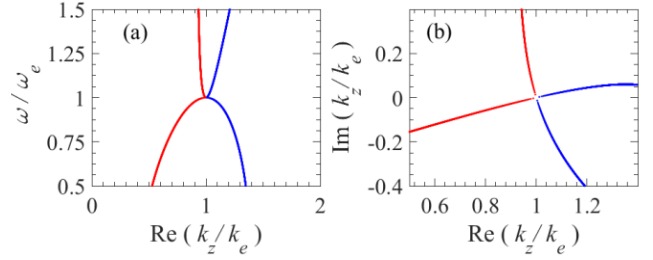

FIG. 7. (a) Dispersion diagram [Re($k_z$) –$\omega$] and (b) complex $k_z$ trajectory varying as a function of frequency of a chain, developing a second order EPD at $\omega_e d/c = 0.1$. Here, we used the same chain's parameters as in Fig. 3, except that the EPD is designed to occur at higher frequency.

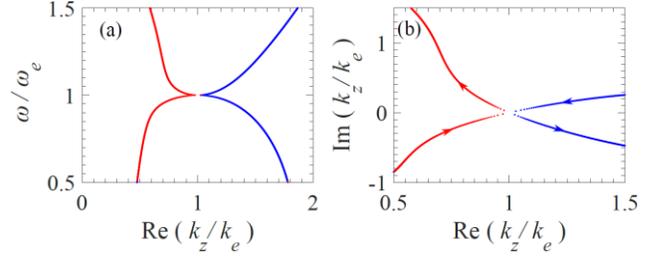

FIG. 8. (a) Dispersion diagram [Re($k_z$) –$\omega$] and (b) complex $k_z$ trajectory varying as a function of frequency of a chain, developing a second order EPD at $\omega_e d/c = 0.5$. Here, we used the same chain's parameters as in Fig. 3, but the EPD frequency is even higher than that in Fig. 7.

$\alpha_2 = (-3.61 + 4.98) \times 10^{-32}$ [$\text{C}\,\text{m}^2\text{V}^{-2}$] for $\omega_e d/c = 0.1$, and

$\alpha_1 = (-3.94 - i6.06) \times 10^{-32}$ [$\text{C}\,\text{m}^2\text{V}^{-2}$],

$\alpha_2 = (-4.19 - i4.74) \times 10^{-32}$ [$\text{C}\,\text{m}^2\text{V}^{-2}$] for $\omega_e d/c = 0.5$ (see table I). We highlight that when the frequency increases the required polarizabilities $\alpha_1$ and $\alpha_2$ to realize an EPD do not satisfy the perfect conjugate condition in (19). In addition, it can be observed from the complex $k_z$ trajectories in Fig. 7(b) and 8(b) that the modes no longer have purely real $k_z$ values for $\omega < \omega_e$, especially for larger $\omega_e d/c$. Nonetheless, the EPD occurrence at $\omega_e$ is evident, even when gain and loss are not symmetric. We point out that near the EPD point in Figs. 7 and 8, in view of the frequency detuning, modes tend to lose symmetry around $k_z = k_e$, meaning that the gain and loss asymmetry causes another form of perturbation near the EPD, especially at high frequencies. In order to capture this asymmetry, one should consider additional terms in the fractional power series expansion (12).

### B. Fourth-Order EPD and Degenerate Band Edge

A fourth-order degeneracy indicates that all four eigenstates of the system (7) coalesce, and this can only be in the form $\zeta_e \equiv e^{ik_e d}$ with $\zeta_e = -1$, i.e., $k_e d = \pi$. This condition occurs

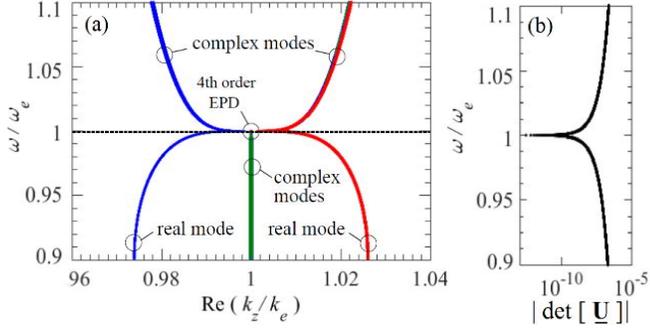

FIG. 9. (a) Dispersion diagram [Re($k_z$) –$\omega$] of the four modes of the chain exhibiting a fourth-order degeneracy at the band edge and (b) the corresponding magnitude of the determinant of similarity **U**. The chain has $d=a=100$ nm with $\omega_e d/c = 0.02$ and $k_e = \pi/d$.

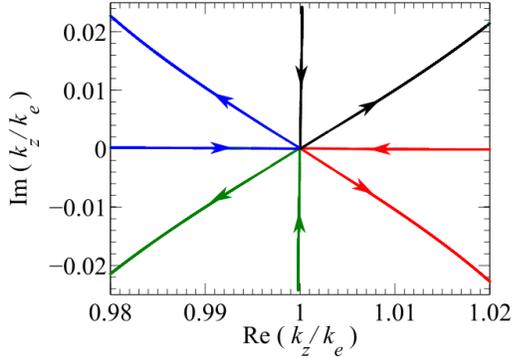

FIG. 10. Complex $k_z$ trajectory plane varying as a function of frequency with the four modes coalescing at $k_e$. The chain has the same parameters used in Fig. 9.

in the middle of the band-edge of the periodic structure's Brillouin zone $k_z \in [0, 2\pi/d]$. For lossless structures, this condition has been conventionally referred to as DBE. Typical examples of DBE effects have been shown in lossless photonic crystals as in [12,13,16,17] and other waveguiding structures [24,25]. Here, however, we show for the first time that the chain develops this fourth-order EPD thanks to the gain-loss interplay, and taking advantage of the natural mode coalescence at the band edge. For the same parameters of the chain discussed in Section II.A, we select $\omega_e d/c = 0.02$ and $k_e d = \pi$. Under this condition, an EPD is attained for $\alpha_1^e \cong (\alpha_2^e)^* = (5.2 - i6.78) \times 10^{-31}$ [Cm$^2$V$^{-2}$] (see table II for $\alpha_1^e$ and $\alpha_2^e$ up to 6 significant digits). The corresponding dispersion relation is shown in Fig. 9(a). Once again, also shown [in Fig. 9(b)] is $|\det(\mathbf{U})|$, which vanishes at $\omega_e$. Similar to the second-order EPD example in Fig. 4, the period is deeply subwavelength ($d \approx 0.003\lambda_e$) and the EPD condition

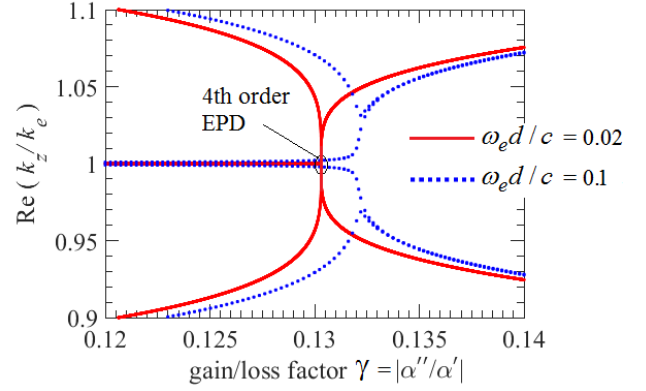

FIG. 11. Positive branch of Re[$k_z$] varying as a function of the gain/loss parameter $g = |\alpha''/\alpha'|$, demonstrating the detuning near a 4$^{th}$ order EPD for two cases of the normalized frequency $\omega_e d/c$. Chain parameters are as in Fig. 9 and $\alpha' = \text{Re}(\alpha_1^e)$ is taken from table II for each case. Note that the perfect gain/loss symmetry condition does not provide a clear EPD for the case with higher frequency.

corresponds to the perfect gain and loss balance and conjugate symmetry condition (19).

By letting $\delta \equiv (\omega - \omega_e)$ in (14), and retaining the first term, the asymptotic dispersion relation near such fourth-order degeneracy is given by $(k_z - k_e)^4 \cong h^4 \delta$, with $h^4 = 24 \partial^4 k_z / \partial \omega^4$ at $\omega = \omega_e$ and $k_z = k_e$. Figure 10 shows the complex $k_z$ trajectory, as a function of frequency. We observe branches of purely real $k_z$ modes and two branches of complex conjugates ones, for $\omega < \omega_e$, coalescing at the EPD for $\omega = \omega_e$, and then evolving into four complex modes for $\omega > \omega_e$.

We also illustrate in Fig. 11 the detuning from the fourth-order EPD by varying the gain and loss parameter $\gamma$ defined in Section III.A, i.e., having conjugate-symmetry in the polarizabilities, $\alpha_{1,2} = \alpha'(1 \pm i\gamma)$. More specifically, we show the positive branches of Re($k_z$) for two different normalized frequencies $\omega_e d/c$, with the corresponding $\alpha' = \text{Re}(\alpha_1^e)$ taken from Table II for each case.

Analogous to the second-order EPD, the occurrence of the fourth-order EPD with perfectly balanced gain/loss and complex-conjugate polarizabilities (i.e., $\mathcal{PT}$-symmetric) is evident for the smaller electrical period ($\omega_e d/c = 0.02$), but it does not hold for the case with $\omega_e d/c = 0.1$. Once again, large values of $\omega_e d/c$ imply that the dispersion relation is deformed in the vicinity of the gain/loss balance condition at which an EPD is expected ($\gamma$

≈ 0.132), and the EPD is no longer observable (as seen in Fig. 11 for the case with $\omega_e d/c = 0.1$).

For better illustration, we show in Figs. 12 and 13 the dispersion relationships and the complex $k_z$ trajectories near the fourth-order EPD for $\omega_e d/c = 0.1$ and $\omega_e d/c = 0.5$, respectively. Once again, the occurrence of a fourth-order EPD is possible, without requiring conjugate symmetry of the polarizabilities. The corresponding polarizabilities of the chain are $\alpha_1^e = (5.225 - i\,0.686) \times 10^{-32}$ [Cm$^2$V$^{-2}$], $\alpha_2^e = (0.522 + i0.691) \times 10^{-32}$ [Cm$^2$V$^{-2}$] for $\omega_e d/c = 0.1$, and $\alpha_1^e = (6.0517 - i4.4607) \times 10^{-32}$ [Cm$^2$V$^{-2}$], $\alpha_2^e = (5.37 + i0.12) \times 10^{-32}$ [Cm$^2$V$^{-2}$] for $\omega_e d/c = 0.5$ (see table II). Results in Fig. 12 and 13 show that the wavenumber trajectory around the fourth-order EPD frequency behaves differently compared to the case in Fig. 9, suggesting that one should consider a larger number of terms in (14) to approximate the eigenstate characteristics (eigenvalue and eigenvectors) near the EPD frequency. The same conclusion applies when other structural parameters are detuned. Nevertheless, by proper tuning of the polarizabilities, one can still attain the remarkable features of fourth-order EPDs, in terms of a high Q factor due to a dramatic reduction group velocity.

TABLE II. REQUIRED CHAIN POLARIZABILITIES TO REALIZE A FOURTH ORDER EPD AT DIFFERENT NORMALIZED FREQUENCIES (UP TO 6 SIGNIFICANT DIGITS).

| $\omega_e d/c$ | $\alpha_1^e \times 10^{-32}$ [Cm$^2$V$^{-2}$] | $\alpha_2^e \times 10^{-32}$ [Cm$^2$V$^{-2}$] |
|---|---|---|
| 0.02 | 5.20205 − i0.677854 | 5.20201 + i0.677894 |
| 0.1 | 5.22558 − i0.686084 | 5.22077 + i0.691123 |
| 0.5 | 6.05176 − i0.446073 | 5.37141 + i1.18834 |

As a final remark, it is worth highlighting that the perturbation of the fourth-order EPD eigenstate with frequency or imbalance of gain and loss is much stronger than the second-order counterpart, since the perturbation factor $\delta$ in the fractional expansion in (12) and (14) dictates that $|\delta|^{1/4} > |\delta|^{1/2} > |\delta|$ for $|\delta| \ll 1$. Accordingly, a small structural perturbation can lead to a significant measurable modification of the spectral evolution of the system near these EPDs, leading to strongly enhanced sensitivity. This can find important applications to sensing [74].

## V. CONCLUSION AND DISCUSSION

We have demonstrated the occurrence of EPDs of second and fourth order in coupled linear chains of scatterers with properly tuned gain and loss in the dipolar polarizabilities. We also have elucidated possible connections with the $\mathcal{PT}$-symmetry concept. In addition, we have demonstrated the impact of gain and loss imbalance as well as conjugate asymmetry of the scatterers' polarizabilities on both second- and fourth-order EPDs. Our results rely on a TB-based approach formulated in terms of transfer matrix. We highlight that the TB approach is an approximation of the more accurate fully-periodic GF method [61,62]. Nonetheless, we have observed good agreement between the TB approach and the fully-periodic GF in analyzing the scattering properties of finite-chains with balanced gain and loss near a long-wavelength EPD, as typical in photonic bandgap structure analyses [52]. These aspects will be investigated more in depth in future studies.

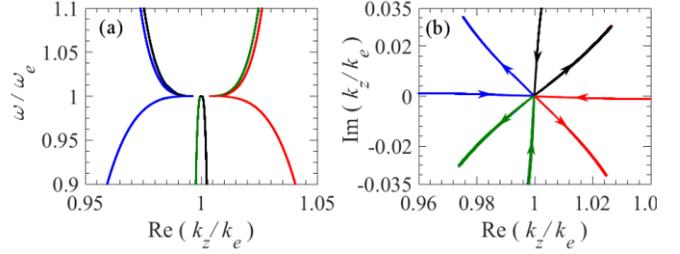

FIG. 12. (a) Dispersion diagram [Re($k_z$) −ω] and (b) complex $k_z$ trajectory varying as a function of frequency of a chain, developing a fourth order EPD at $\omega_e d/c = 0.1$. The chain has the same parameters used in Fig. 9.

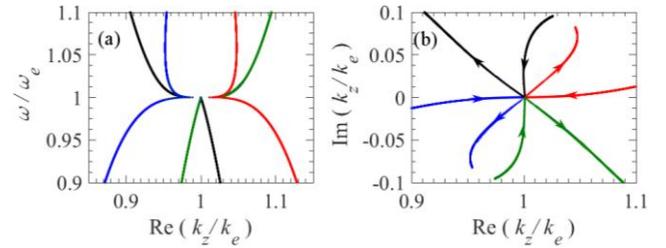

FIG. 13. (a) Dispersion diagram [Re($k_z$) −ω] and (b) complex $k_z$ trajectory varying as a function of frequency, for a chain with the same the same geometrical parameters used in Fig. 9. It develops a fourth order EPD at $\omega_e d/c = 0.5$ when the perfect gain/balance condition does not hold anymore.

Our analysis provides some new insights into how EPDs can manifest in general discrete coupled mode structures. These properties can also be harnessed for sensing applications, enhancing non-linear effects (including second harmonic generation and unprecedented soliton propagation), as well as lowering the threshold for lasing (as demonstrated in [16] for lossless DBE structures with extrinsic gain). Moreover, the structures of interest can be implemented by using plasmonic particles for applications ranging from near-field enhancement to super-resolution at optical wavelengths.

**ACKNOWLEDGEMENT**

The work of M. O. and F. C. is supported by Air Force Office of Scientific Research under Grant FA9550-15-1-0280, and also by MURI Grant FA9550-12-1-0489 administered through the University of New Mexico.